\documentclass[conference]{IEEEtran}
\IEEEoverridecommandlockouts
\usepackage{cite}
\usepackage{amsmath,amssymb,amsfonts}
\usepackage{algorithmic}
\usepackage{graphicx}
\usepackage{listings}
\usepackage{textcomp}
\usepackage{xcolor}
\usepackage{url}
\usepackage{xcolor} 
\usepackage{algorithm2e} 
\usepackage{braket} 
\usepackage{url}

\newcommand{\new}[1]{\textcolor{black}{#1}}

\usepackage{xcolor}
\usepackage{tcolorbox}
\usepackage{balance}
\usepackage{hyperref}

\definecolor{codegreen}{rgb}{0,0.6,0}
\definecolor{codegray}{rgb}{0.5,0.5,0.5}
\definecolor{codepurple}{rgb}{0.58,0,0.82}
\definecolor{backcolour}{rgb}{0.95,0.95,0.92}

\lstdefinestyle{mystyle}{
    commentstyle=\color{codegree.n},
    keywordstyle=\color{magenta},
    numberstyle=\tiny\color{codegray},
    stringstyle=\color{codepurple},
    basicstyle=\ttfamily\footnotesize,
    breaklines=true,
    keepspaces=true,
    tabsize=1,
}

\def\BibTeX{{\rm B\kern-.05em{\sc i\kern-.025em b}\kern-.08em
    T\kern-.1667em\lower.7ex\hbox{E}\kern-.125emX}}

\begin{document}

\title{ProvideQ: A Quantum Optimization Toolbox
\thanks{
This work was supported by the German Federal Ministry for Economic Affairs and Energy (ProvideQ, 01MQ22006D) and the German Federal Ministry of Research, Technology and Space (QuSol, 13N17170).} 
}

\author{
\IEEEauthorblockN{
Domenik Eichhorn\textsuperscript{1}, 
Nick Poser\textsuperscript{1}, 
Maximilian Schweikart\textsuperscript{1,2}, 
Ina Schaefer\textsuperscript{1}
}

\IEEEauthorblockA{
\textsuperscript{1}Karlsruhe Institute of Technology, Germany, 
\{domenik.eichhorn, nick.poser, ina.schaefer\}@kit.edu
}

\IEEEauthorblockA{
\textsuperscript{2}University of Oxford, United Kingdom, 
maximilian.schweikart@cs.ox.ac.uk
}
}

\maketitle

\begin{abstract}
Hybrid solvers for combinatorial optimization problems combine the advantages of classical and quantum computing to overcome difficult computational challenges. Although their theoretical performance seems promising, their practical applicability is challenging due to the lack of a technological stack that can seamlessly integrate quantum solutions with existing classical optimization frameworks. We tackle this challenge by introducing the ProvideQ toolbox, a software tool that enables users to easily adapt and configure hybrid solvers via Meta-Solver strategies. A Meta-Solver strategy implements decomposition techniques, which splits problems into classical and quantum subroutines. The ProvideQ toolbox enables the interactive creation of such decompositions via a Meta-Solver configuration tool. It combines well-established classical optimization techniques with quantum circuits that are seamlessly executable on multiple backends. This paper introduces the technical details of the ProvideQ toolbox, explains its architecture, and demonstrates possible applications for several real-world use cases. Our proof of concept shows that Meta-Solver strategies already enable the application of quantum subroutines today, however, more sophisticated hardware is required to make their performance competitive.
\end{abstract}


\section{Introduction}

Current roadmaps on established quantum technologies, such as superconducting qubits, trapped ions, or photonics, claim optimistic assumptions of medium-scale Noisy Intermediate Scale Quantum (NISQ) being available within the next years~\cite{byrd2023quantum}. 
Given these recent advancements, quantum algorithms such as QAOA~\cite{farhi2014quantum} and VQE~\cite{peruzzo2014variational} could become competitive in solving optimization problems and outperform classical alternatives for specific use cases~\cite{shaydulin2024evidence}.
Coupling these potential quantum advantages with highly optimized classical algorithms is a promising approach for solving currently intractable large-scale problems.
The development of hybrid quantum-classical solver technologies has the potential to greatly improve the efficiency of modern industrial processes~\cite{perez2024quantum}, but their practical application is currently impeded by many open challenges.

Although the availability of scalable quantum hardware is currently the most critical factor for quantum speedups, another as important aspect is missing: a hybrid software stack~\cite{10234290}.
To fully utilize the advantages of hybrid optimization algorithms, the interaction between classical and quantum subroutines needs to be orchestrated efficiently.
To truly leverage the advantages of quantum computers, we need to decompose classical optimization models and algorithms and analyze which scenarios could benefit from the application of quantum computing.
This intend is ambitious and requires well-optimized software tools to efficiently connect the classical and quantum realms. 
Existing quantum software solutions for combinatorial optimization, such as Qiskit-Optimization~\cite{Qiskit} or Qrisp~\cite{seidel2024qrisp}, require developers to (a) understand and implement/adapt the details of quantum algorithms, (b) implement their own error mitigation techniques to deal with the noise of NISQ devices, and (c) optimize their circuits for the underlying hardware.
There is a lack of software tools that can completely hide the complexity of these three steps behind well-designed interfaces, making the adaptation of hybrid optimization techniques infeasible for non-experts.

In this paper, we address the challenge of a lacking hybrid quantum-classical software stack by introducing a new tool that facilitates convenient access to hybrid algorithms for combinatorial optimization problems.
We introduce the \textit{ProvideQ toolbox}, which seamlessly integrates state-of-the-art classical optimization techniques with quantum subroutines, thereby enabling easy adaptation of quantum solutions without losing the benefits of existing classical techniques.  
The primary objective of the ProvideQ toolbox is to facilitate a novel method, designated as \textit{Meta-Solving}~\cite{eichhorn2024hybridmetasolvingpracticalquantum}. 
The Meta-Solving process supports the user by providing technologies to decompose large optimization models into multiple subproblems via so-called Meta-Solver strategies and then providing configurable hybrid quantum-classical solvers to solve them. 
The ProvideQ toolbox can be used to easily experiment with different solving strategies without having to worry about the complexity of the integrated quantum aspects. 
Users can therefore focus on different aspects of the problem-solving process: such as the creation of better models, the combination of different solving strategies, and the optimization of parameters.

Our current prototype incorporates established standards, parsers, solvers, and decompositions for a range of well-known algorithmic problems. 
We support problems such as vehicle routing, traveling salesperson, knapsack, max-cut, boolean satisfiability (SAT), and quadratic unconstrained binary optimization (QUBO). Our quantum solvers are currently based on QAOA~\cite{farhi2014quantum} or Grovers algorithm~\cite{grover1996fast}, while our classical solvers reuse existing state-of-the-art algorithms such as LKH-3~\cite{helsgaun2017extension} for vehicle routing, or Horowitz-Sahni~\cite{horowitz1974computing} for knapsack. 
Quantum algorithms can be simulated locally or sent to backends from established vendors, given that the user has a valid API key.
The ProvideQ toolbox is open source and available on GitHub: 
\href{https://github.com/ProvideQ}{https://github.com/ProvideQ}.

\section{Background}
\label{sec:Background}

The background section presents a concise summary of the state-of-the-art methods used in classical, quantum, and hybrid quantum-classical optimization. 
Additionally, we explain polylithic solvers and the Meta-Solving approach as foundational elements for interactive solver development.

\subsection{Hybrid Optimization}

Many applications require solving combinatorial optimization problems such as Traveling Salesperson (TSP), Vehicle Routing Problem (VRP), or Knapsack on a large scale. 
As the NP-complete character of these problems hinders the creation of general algorithms that can solve them efficiently, new methods are needed to bypass their exponential growth constraints~\cite{hochba1997approximation}.
Current methodologies utilize various techniques that prioritize the search for a satisfactory solution over an optimal one. 
This involves, for example, the use of approximations~\cite{hochba1997approximation} and heuristics~\cite{rardin2001experimental}, with the goal of decreasing computation times while retaining a good solution. 

Since the initial introduction of quantum algorithms such as Grover~\cite{grover1996fast} and Shor~\cite{shor1999polynomial}, it is widely believed that quantum computers have the potential to outperform their classical counterparts by providing up to superpolynomial speedups~\cite{abbas2024challenges}. 
However, we are currently not able to leverage these theoretical speed-ups in practical use cases.
With the limitations of NISQ hardware and the particularities of optimization problems in mind, specialized algorithms such as the Quantum Approximate Optimization Algorithm (QAOA)~\cite{farhi2014quantum}, and the Variational Quantum Eigensolver (VQE)~\cite{peruzzo2014variational} have been designed. 
These approximation algorithms can deal with the noise of currently available quantum hardware and have been shown to produce reasonable results for small-scale problems when optimized correctly~\cite{tate2023bridging, tate2023warm}. 

Hybrid quantum-classical solvers combine the advantages of classical and quantum optimization algorithms in a new type of technology~\cite{GE2022314}. 
They acknowledge the potential of quantum computing, yet continue to utilize the advantages of highly optimized classical solver frameworks.
Applying hybrid quantum-classical technologies requires us to decompose an algorithmic problem into multiple subroutines, have a efficient orchestration between the quantum and classical realms, and have knowledge about the benefits of each technology. 
Depending on the exact algorithm or solver framework, the utilization of classical and quantum subroutines can vary a lot. 
QAOA, for instance, can be considered a hybrid quantum-classical algorithm, as it combines the execution of a parametrized quantum circuit with classical optimizers. Thus, we utilize classical combinatorial optimization technologies to improve the results of the quantum algorithm.
The Quantum-Enhanced Monte Carlo method~\cite{layden2023quantum} is an opposing example, which uses variations of the Grover algorithm to speed up the sampling subroutines of a classical algorithm.
\new{
Hybrid quantum-classical optimization is a rapidly growing field, but a practical inclusion of quantum computing still requires us to take a different look at solving optimization problems.
}

\subsection{Polylithic Modeling and Meta-Solving}

The polylithic modeling approach was introduced by Josef Kallrath~\cite{Kallrath2011}, and describes how the creation of multiple smaller models, instead of one large monolithic model, can benefit the solving of combinatorial optimization problems.
By decomposing the input problem into smaller problems that can be fed into other models, large problems can be tackled with a divide-and-conquer approach.
Examples of polylithic approaches are decomposition techniques such as Benders~\cite{benders2005partitioning} and Dantzig-Wolfe~\cite{dantzig1961decomposition}, or advanced (Mixed) Integer Programming solvers that use resolution strategies combined with branch and cut~\cite{Kallrath2011}.

\begin{figure}[h]
    \centering
    \includegraphics[width=0.9\linewidth]{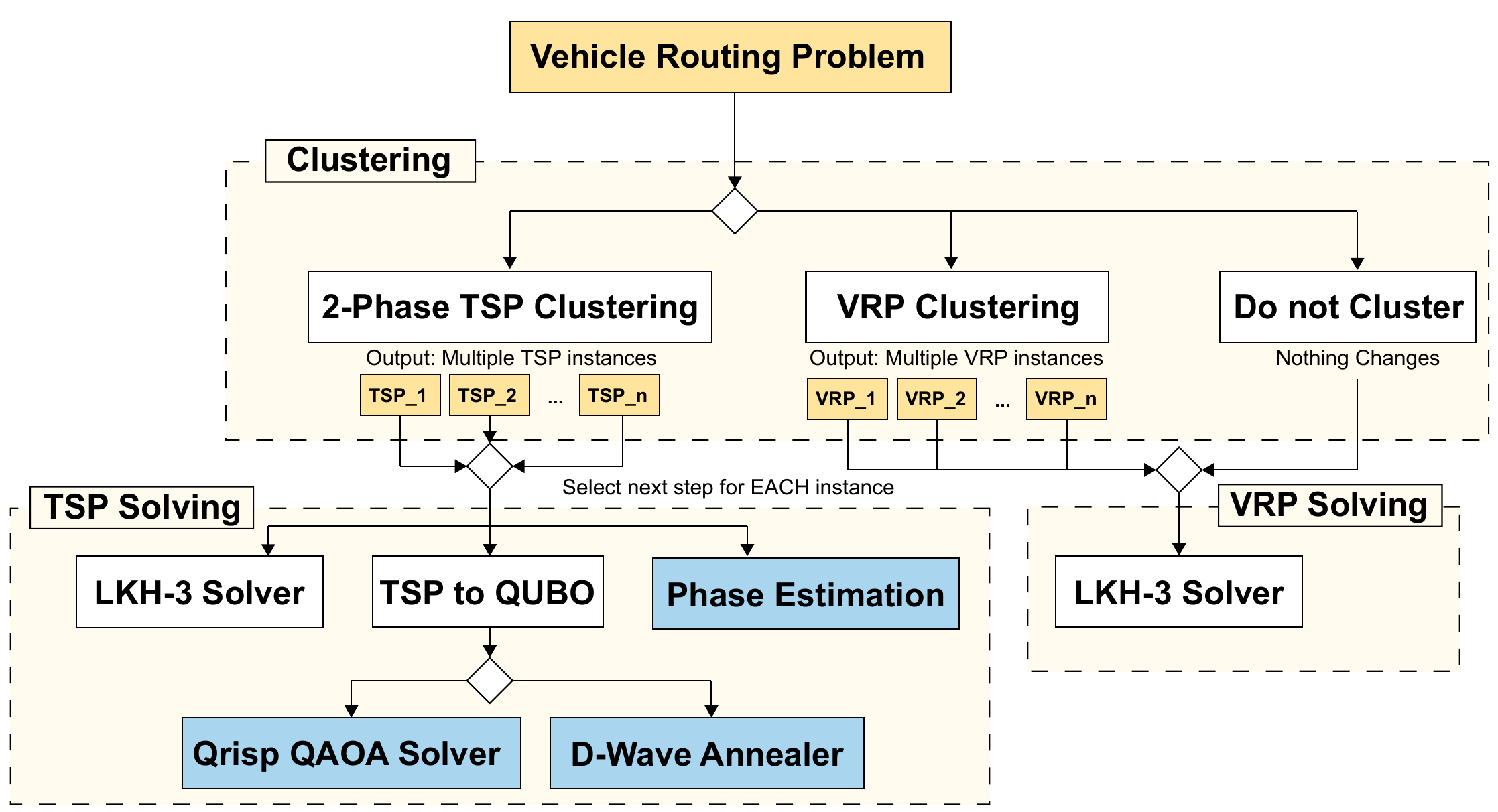}
    \caption{Running Example for a Vehicle Routing Meta-Solver Strategy~\cite{eichhorn2024hybridmetasolvingpracticalquantum}.}
    \label{fig:vrp-running-example}
\end{figure}

The decomposition of large mathematical models is also a necessary step when applying quantum computing methods in the NISQ era. 
To extend the benefits of polylithic modeling with quantum subroutines, we introduced a concept to model hybrid quantum-classical solution processes, which is called Meta-Solving~\cite{eichhorn2024hybridmetasolvingpracticalquantum}.
Meta-Solving adapts the philosophy of creating multiple smaller models and provides techniques that help users with the model decomposition, the solving of the derived submodels, and finally the composition of results. 
We reach this goal by defining so-called Meta-Solver strategies.
Meta-Solver Strategies define multiple approaches to solve an optimization problem, which usually include multiple subroutines and decomposition techniques. 
An example for such a Meta-Solver strategy is shown in Figure~\ref{fig:vrp-running-example}. 
It describes several approaches on how to solve VRP problems, including two different ways to decompose them (the two clustering algorithms), and several ways to solve VRP and TSP submodels. 
The important subroutines for this paper are
the \textit{2-Phase TSP Clustering} proposed by Laporte et al.~\cite{laporte2002classical}, which divides the VRP problem into a clustering phase that can be represented as a knapsack problem, and routing phase that can be represented as a TSP problem~\cite{feld2019hybrid};
the \textit{TSP to QUBO} transformation that is based on the Ising formulations from Lucas~\cite{lucas2014ising};
and the \textit{LKH-3 Solver}~\cite{helsgaun2017extension}, which is a classical state-of-the-art solver for VRP and TSP problems.
The ProvideQ Toolbox is a prototype aiming to show the possibilities of the Meta-Solving approach when adapted in software tools. 
We will continue to use the VRP Meta-Solver Strategy Figure~\ref{fig:vrp-running-example} as a running example to visualize certain aspects of ProvideQ.

\newpage

\section{Using the ProvideQ Toolbox}
\label{sec:using_the_provideq_toolbox}

\textit{Providing Quantum} is an ambitious goal that requires the combination of expert knowledge from the fields of physics, mathematics, and computer science.
For quantum computers to make a meaningful impact to modern industry, medicine, or other fields, they need to become easily accessible to non-expert users.
With the \textit{ProvideQ} toolbox, we present a new tool that aims to reduce the entry barrier to quantum optimization techniques by providing convenient access to new quantum technologies and combining them with the existing classical state-of-the-art.
Designed around the concept of Meta-Solving, our toolbox implements workflows that allow users to configure, execute, and evaluate powerful hybrid solvers. 

\begin{figure}[h]
    \centering
    \includegraphics[width=\linewidth]{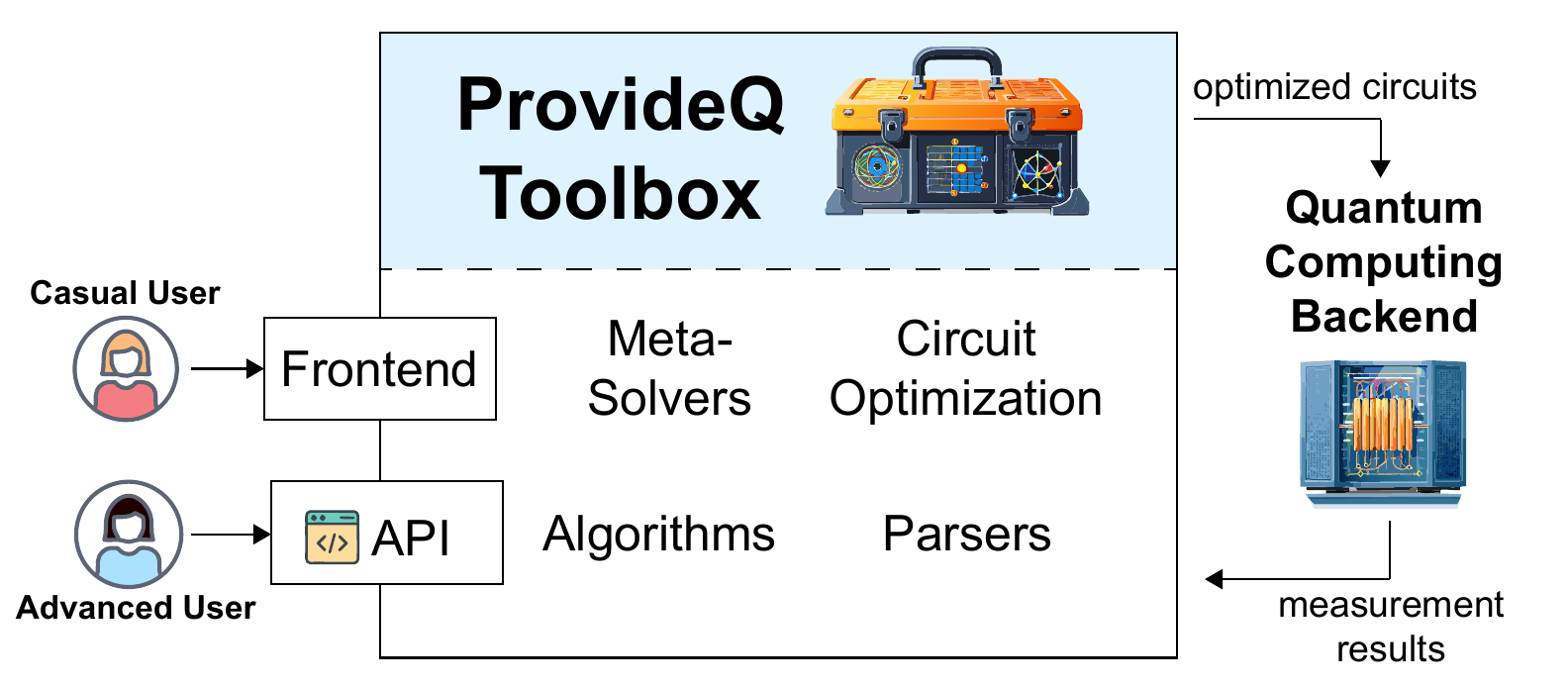}
    \caption{Overview of the ProvideQ Toolbox.}
    \label{fig:user_interaction}
\end{figure}

As visualized in Figure~\ref{fig:user_interaction}, the toolbox implements several Meta-Solvers, Algorithms, Parsers, and Circuit Optimizations.
Users can interact with the ProvideQ toolbox over two different interfaces: a technical \textit{API}, or a user-friendly \textit{Frontend}. 
Both interfaces provide different advantages.
The frontend is designed for a \textit{Casual User} who wants to explore the opportunities of hybrid solvers and test out new solution strategies. 
The API is designed for an \textit{Advanced User} who wants to integrate hybrid solvers into external tools or solve many different types of optimization problems in batches. 
A detailed user journey that exemplifies how a user can interact with the frontend is described in Section~\ref{sec:frontend_journey}, a user journey for the API is described in Section~\ref{sec:api_journey}.

\subsection{Solver Configuration with a Graphical User Interface}
\label{sec:frontend_journey}

\begin{figure}[h]
    \centering
    \includegraphics[width=\linewidth]{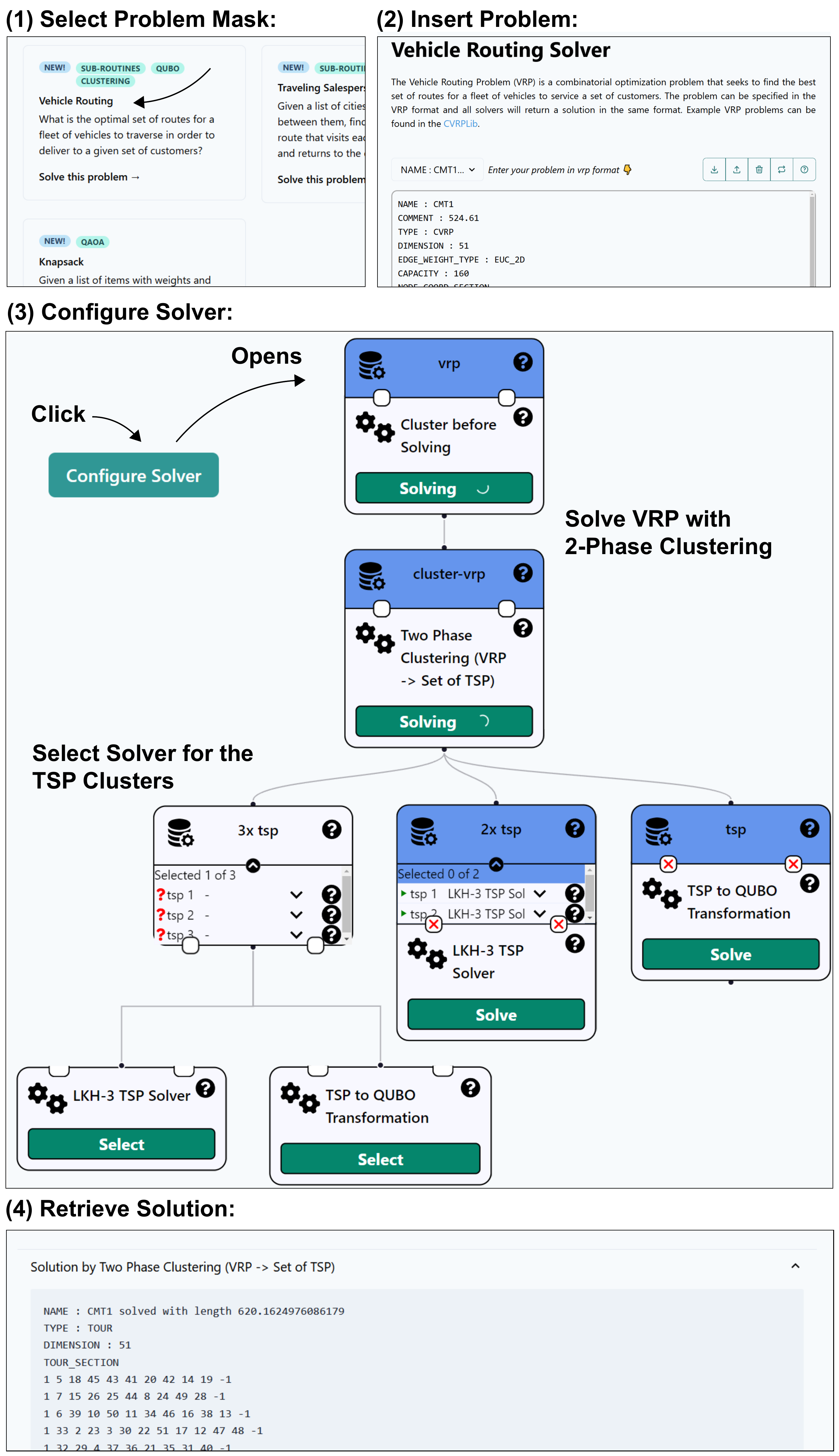}
    \caption{Example Usage of the ProvideQ Frontend.}
    \label{fig:frontend_usage}
\end{figure}

The \textit{ProvideQ Frontend} provides an easy-to-use interface that allows users to explore different solution strategies without having to understand the technical details. 
As visualized in Figure~\ref{fig:frontend_usage}, a user is welcomed on a landing page (1) on which a \textit{Problem Mask} can be selected. 
These include, for instance, the VRP, TSP, and knapsack problem. Each Problem Mask provides a description of the problem, a standardized input field, and basic utilities, such as the ability to load example problems.
In our example, the user selects the Vehicle Routing Problem Mask (2).
The user inserts a VRP instance and then clicks on the \textit{Configure Solver} button, which will open the \textit{Solver Configurator} (3).
The Solver Configurator presents the user with a visualized Meta-Solver strategy, guiding the user through the solver configuration process by proposing possible combinations of subroutines.

In the example shown in step (3), the user decided to solve the problem with the 2-Phase TSP Clustering approach from our running example (see Figure~\ref{fig:vrp-running-example}), which includes a clustering and a routing part.
The clustering part has already been solved and has produced a set of six TSP problems, for which the user is currently selecting the concrete solvers.
One TSP problem should be solved by a QUBO transformation, two with LKH-3, and for the other three problems the solver is still unselected.
When clicking on solve, the LKH-3 solver will calculate a result for the TSP, whereas the QUBO transformation will transform the TSP into a QUBO.
The user is then required to continue the solver configuration by selecting a concrete QUBO solver.
Once the solver configuration is finalized, the ProvideQ toolbox combines the subroutine solutions to return a solution to the original VRP problem (4). 

\subsection{Solver Configuration with Programming Interfaces}
\label{sec:api_journey}

The ProvideQ Frontend visualizes Meta-Solver strategies in a user-friendly manner and enables interested users to test hybrid quantum-classical solvers and the capabilities of the toolbox.
However, the frontend is not suitable for more advanced use-cases such as batch processing or the integration as a component into larger pieces of external software.
For these use-cases we offer a full Application Programming Interface (API) that can be accessed directly via HTTPS requests or via a Python wrapper\footnote{ 
The HTTPS API can be accessed via 
~\href{https://api.provideq.kit.edu/}{https://api.provideq.kit.edu/}, 
or self-hosted from
~\href{https://github.com/ProvideQ/toolbox-server}{https://github.com/ProvideQ/toolbox-server}.
The Python wrapper is available at
~\href{https://github.com/ProvideQ/toolbox-python-api}{https://github.com/ProvideQ/toolbox-python-api}.}.
To solve our VRP running example via the HTTPS API, a user starts by making a POST request:

\begin{tcolorbox}[colback=black!5, colframe=black!50, sharp corners, boxrule=0.2mm]
\small
\ttfamily
\textbf{POST} /problems/cluster-vrp \{ \\
  "typeId": "cluster-vrp", \\
  "input": "[the VRP problem]" \}
\end{tcolorbox}

This request will create the problem instance on the server side and return a problem id to the user. 
Now, the solver for the problem needs to be configured. 
To be consistent with the running example and the solver from Section~\ref{sec:frontend_journey}, we set the Two-Phase Clustering as the VRP solver. 
This is done by sending a PATCH request to the toolbox:

\begin{tcolorbox}[colback=black!5, colframe=black!50, sharp corners, boxrule=0.2mm]
\small
\ttfamily
\textbf{PATCH} /problems/cluster-vrp/\{problemId\} \{
  "state": "SOLVING", \\
  "solverId": "edu.kit.provideq.toolbox.\\
        vrp.clusterer.TwoPhaseClusterer" \}
\end{tcolorbox}

One part of the shown PATCH request is setting the state of the problem to \textit{SOLVING}, which will initiate the solving process on the server side.
In our example, the Two Phase Clustering will decompose the VRP into six TSP clusters and therefore return a list of six new sub-problem ids.
The user can now use these sub-problem ids to configure a solver for each TSP instance via similar PATCH requests.  
Once all six TSP problems have been solved, the subroutines of the Two Phase Clustering will be marked as \textit{SOLVED}, and again, a solution for the original VRP problem is returned.

\section{Technical Realization}

After the general overview presented in Section~\ref{sec:using_the_provideq_toolbox}, we will now introduce the technical details of the ProvideQ toolbox.

\subsection{Architecture Overview}

\begin{figure}
    \centering
    \includegraphics[width=\linewidth]{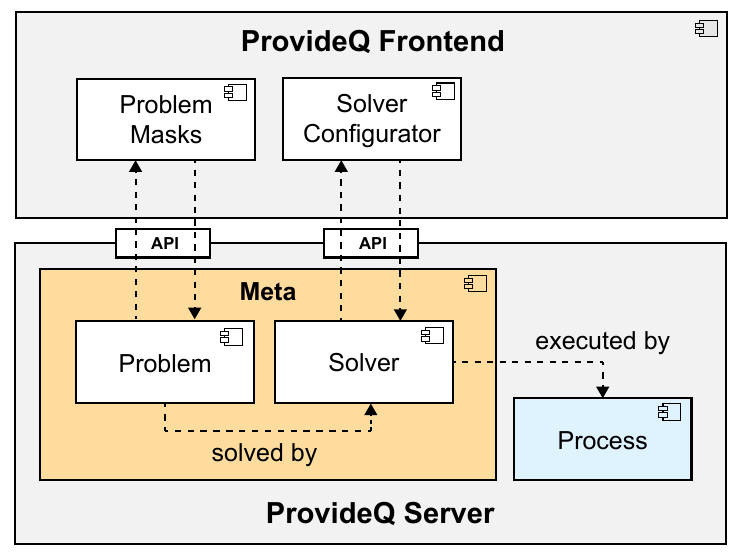}
    \caption{
        Client-Server software architecture of the ProvideQ toolbox. All following figures will reuse the color pattern of this figure, using a yellow color for artifacts associated with the \textit{Meta} component, and a light blue color for artifacts associated with the \textit{Process} component.
    }
    \label{fig:component-diagram}
\end{figure}

To cover the requirements of different types of users, the ProvideQ toolbox is based on a Client-Server architecture, visualized in Figure~\ref{fig:component-diagram}.
The \textit{ProvideQ Server} component acts as the heart of our tool.
It consists of a \textit{Meta} component that realizes Meta-Solver strategies by orchestrating the \textit{Problem} and \textit{Solver} components. 
The architecture ensures low coupling between different solvers and provides abstraction layers that allow a solver to reuse other solvers for subroutines.

The \textit{Problem} component handles the definition of solvable combinatorial optimization problems. 
It defines standards for input and output formats, contains descriptions of all problems, and provides other utilities, such as tools for the estimation of upper and lower bounds.
To solve a problem, we utilize the \textit{Solver} component, which implements concrete algorithms (such as LKH3 or QAOA), decomposition techniques (such as the 2-Phase Clustering), and conversion tools to parse input and output formats into the correct standards. 
The execution of a solver is done by the \textit{Process} component, which handles the networking aspects between different technologies. These include, for instance, the communication with different quantum backends, but also the connection of multiple classical technologies, like calling solvers that are implemented in different programming languages. 
Most parts of the \textit{ProvideQ Server} are implemented in Java using the well-established Spring-Boot framework. 
Exceptions are all performance critical components, such as concrete solvers, special decomposers, or anything connected to quantum circuit deployment.
The technologies behind these parts differ and include C, Rust, and Python code. 
Reason for this variation is that our problem solvers reuse existing implementations of open-source solvers like LKH-3~\cite{helsgaun2017extension} and quantum frameworks like Qiskit~\cite{Qiskit} or Qrisp~\cite{seidel2024qrisp}.

External components can access the \textit{ProvideQ Server} via an API, making it easy to integrate its features into external software.
One such external software is the \textit{ProvideQ Frontend}, which implements the graphical user interface that was previously shown in Figure~\ref{fig:frontend_usage}.
The \textit{ProvideQ Frontend} visualizes all aspects of the \textit{ProvideQ Server} via web pages that are implemented in React.
It consists of a \textit{Problem Mask} component that lets users select a combinatorial optimization problem and provide an input for it, and a \textit{Solver Configurator} that visualizes Meta-Solver strategies.

\subsection{Implemementing Meta-Solver Strategies}
\label{sec:solver-modularization}

\begin{figure}[ht]
    \centering
    \includegraphics[width=\linewidth]{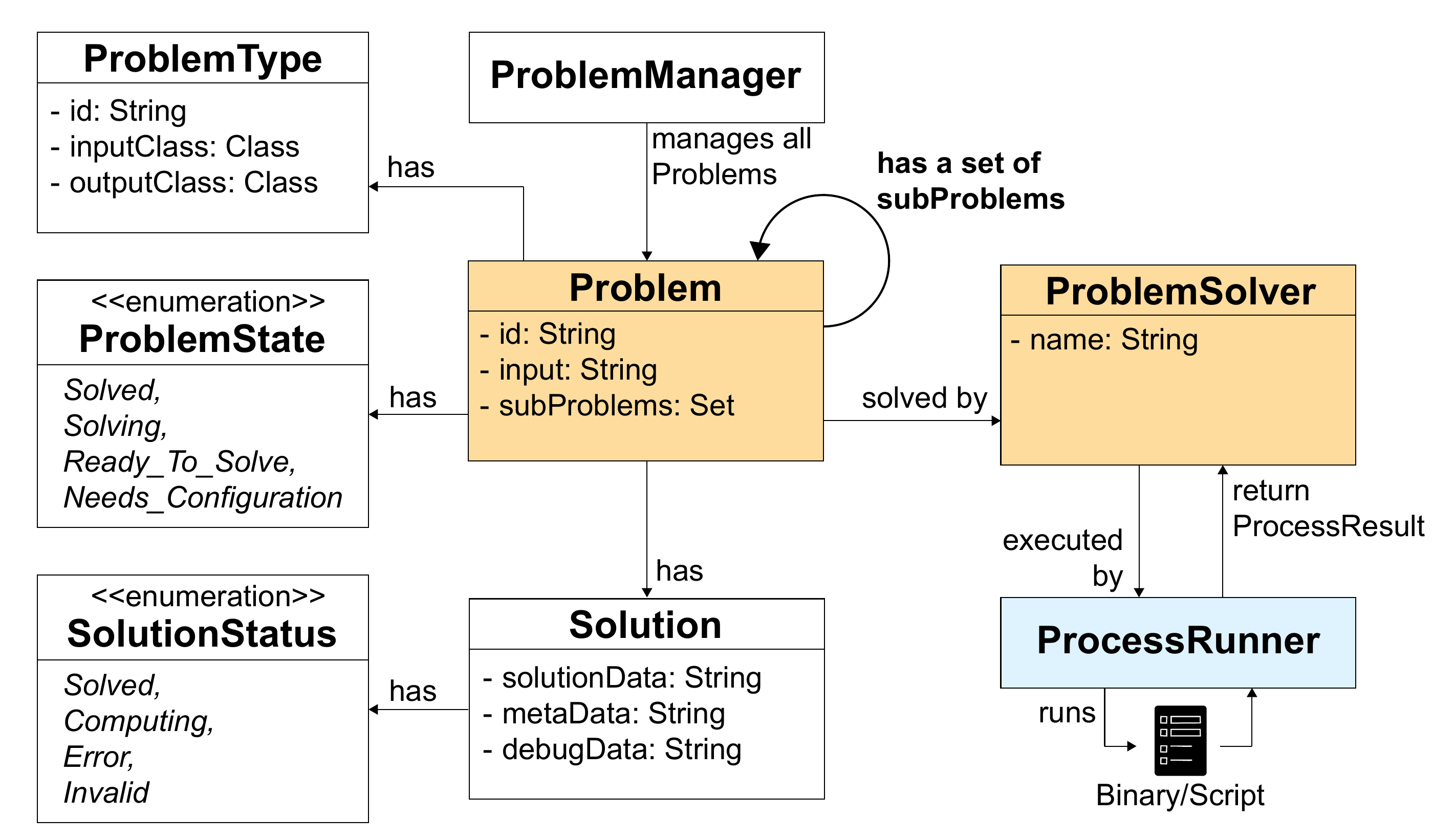}
    \caption{Class Diagram visualizing the architecture of the \textit{Meta} component.}
    \label{fig:solver-modularization}
\end{figure}

We implement Meta-Solver strategies by modularizing solvers, thereby enabling them to function as interchangeable modules within the solving process. 
As shown in Figure~\ref{fig:solver-modularization}, our Meta-Solver architecture revolves around a \textit{Problem} class. 
Each problem object has a unique \textit{id}, an \textit{input}, a \textit{problemType}, a \textit{problemState}, and a \textit{solution}.
The \textit{input} is a string following a standardized formulation of the problem, for example, a model of a combinatorial optimization problem.
The \textit{solution} is set once the problem is solved and contains a result for the problem and some additional meta- and debug data.

The status of problems and their solutions vary during their lifecycle.
Upon initial creation, a problem is set to the \textit{Needs Configuration} state, and its solution is set to null. 
A problem becomes \textit{Ready to Solve} when a \textit{ProblemSolver} is assigned to it.
A \textit{ProblemSolver} is an abstract wrapper around concrete solvers, and is executed by a \textit{ProcessRunner}. Each \textit{problemSolver} object has a \textit{name}, a \textit{description}, a \textit{problemType} and a set of \textit{settings}. 
The \textit{name} and \textit{description} are identifiers that are passed to the API and later utilized to distinguish different problemSolvers.
The \textit{problemType} describes the combinatorial optimization problem that is associated with the \textit{problemSolver} and has to match the \textit{problemType} of the \textit{problem} object. 
The \textit{settings} describe possible configuration options of the \textit{problemSolver}, such as fine-tunable parameters or API tokens for quantum backends.

Once a \textit{problemSolver} is selected and all its settings are configured, the resolution of the \textit{problem} can be initiated by modifying its status to \textit{Solving}. 
The \textit{problemSolver} will now be executed by a \textit{processRunner}, which will consequently create a \textit{solution} object in the \textit{Computing} state. 
Once a result is computed, the \textit{problemState} and is set to \textit{Solved}, and the \textit{solutionStatus} is set to \textit{Solved}, \textit{Error}, or \textit{Invalid}, depending on the result of the \textit{processRunner}. 

A key principle that lifts our approach from a monolithic solver to a polylithic Meta-Solver is the incorporation of the \textit{ProblemManager} and the set of \textit{subProblems} that can be defined by each Problem. 
The construction of subproblems enables the implementation of mathematical decompositions, splitting one large problem into a set of smaller, more manageable units.
Rather than attempting to solve the entire problem at once, we can instead provide the solution to one subroutine and then define a set of additional subproblems that must be solved in order to retrieve an answer to the underlying mathematical challenge. 

The role of the \textit{ProblemManager} is to monitor the creation and progression of all problems and to facilitate their resolution. 
The system guarantees that problems are solved only by solvers whose profiles align with the specified \textit{ProblemType}, and it ensures that the subproblems are executed in the correct sequence. 
In addition, it monitors the status of a problem and notifies other components about their respective lifecycle. 
\subsection{Implementing the Solver Execution}

Most of the solvers included in the ProvideQ toolbox are not implemented by ourselves because we try to reuse well-established existing solvers, such as LKH-3~\cite{helsgaun2017extension}, or the QAOA implementation from Qrisp~\cite{seidel2024qrisp}. 
We include these external pieces of software via our \textit{ProcessRunner} wrapper in the \textit{Process} component. 
This wrapper executes the scripts, binaries, or other artifacts that represent the concrete solver implementation and parses their output back to our internal framework. 
Depending on the particularities of the solver, the execution needs to utilize different types of hardware.

Classical algorithms exhibit special resource utilization patterns with respect to (V)RAM, CPU, and GPU power.
Quantum algorithms do the same thing, but with different kinds of resource, such as the number of required qubits and the circuit depth.
However, compared to their classical counterparts, quantum computers strongly differ in their technological details, which has a high impact on the capability to execute specific circuits.
Thus, when executing quantum circuits, we need to consider a more complex landscape of system details, such as the availability of specific gates, entanglement maps, and error rates. 
Furthermore, quantum computers are not yet widely available. Currently, the execution of quantum circuits can only be achieved through the use of simulators or by accessing quantum backends via cloud computing. 
As a result of all these characteristics, quantum algorithms are generally much more difficult to execute efficiently than classical algorithms.
The ProvideQ toolbox handles this complexity inside the \textit{Process} component, the workflow of which is illustrated in Figure~\ref{fig:deployment}.
The main task of this component is to match algorithms with fitting backends and to deploy them for execution.
The initial decision in the deployment process is whether the subroutine must be executed on a classical or a quantum backend. 

The methodology for handling classical subroutines is straightforward.
The subroutines are executed on the computer on which the ProvideQ Server is hosted, and thus the available resources are dependent on the host machine. 
In contrast, remote execution is mandatory for non-simulated quantum subroutines, given that current quantum computers are always external machines, currently usually hosted by larger vendors. 

The execution of quantum subroutines is centered around quantum circuits. 
Once a quantum solving method (such as QAOA, VQE, Grover, etc.) has been selected, the classical problem is transformed into a quantum circuit. 
The concrete layout of the quantum circuit depends on the specified quantum solving method and the characteristics of the problem instance.
After deriving a quantum circuit, the problem can be processed by interacting with it in two possible ways:
(1) the circuit can be tuned through the application of a circuit optimization technique, or 
(2) the circuit can be transpiled and executed on a selected quantum backend. The execution returns measurements of the quantum systems that are then interpreted as a solution to the problem.

\subsubsection*{(1) Circuit Optimization}
\begin{figure}
    \centering
    \includegraphics[width=0.94\linewidth]{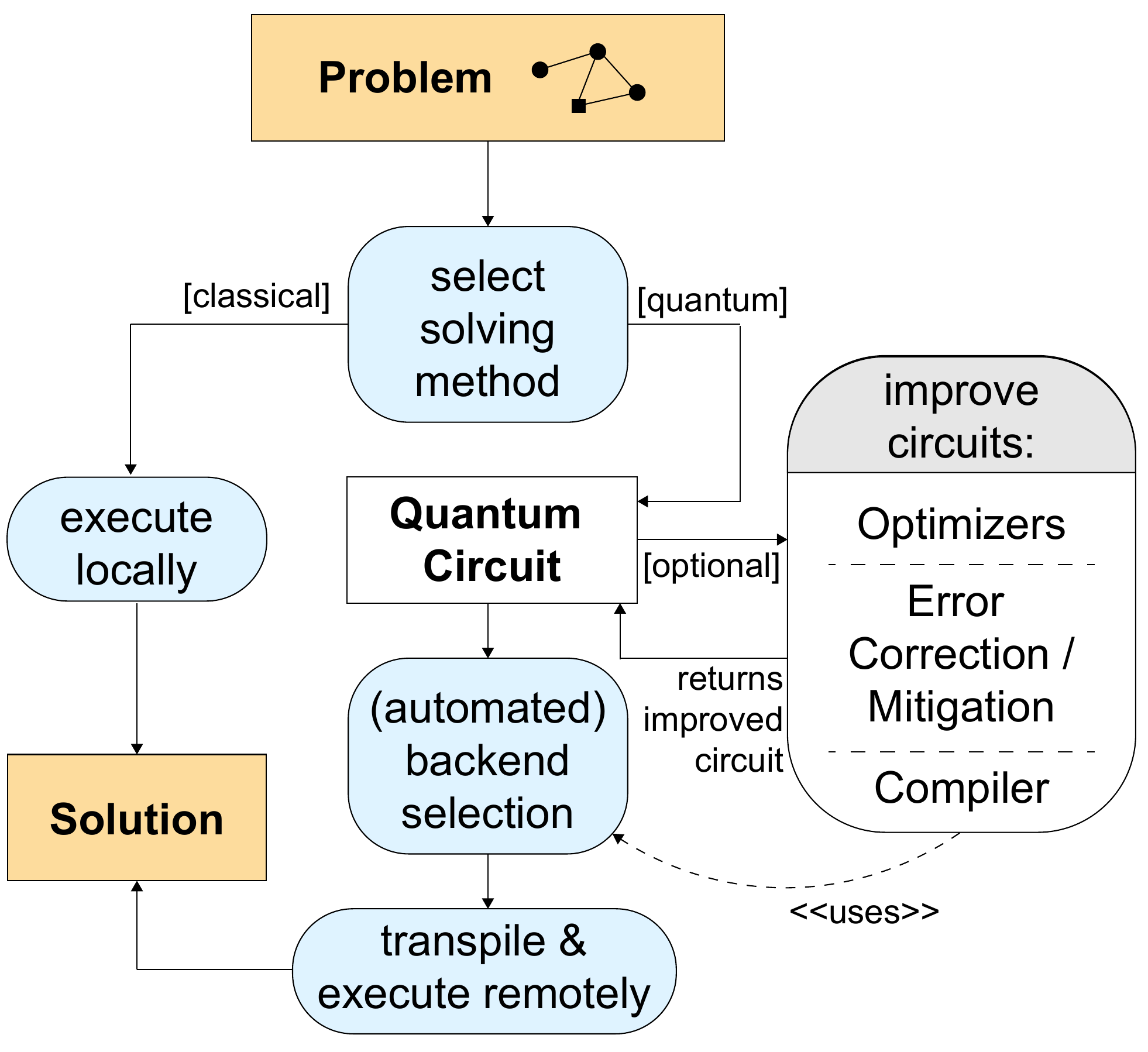}
    \caption{Workflow of the \textit{Process} component.}
    \label{fig:deployment}
\end{figure}

Given the limitations of the currently available NISQ hardware, circuit optimization is essential to achieve competitive results.
As illustrated in Figure~\ref{fig:deployment}, our \textit{Process} component implements a set of optional circuit optimization techniques that can be applied.
Improving quantum circuits is a complex task, and the possible benefit of a circuit optimization technique strongly depends on the selected solving method and the quantum backend.
Although there are known best practices, we usually do not know the best way to optimize a circuit beforehand.
The ProvideQ toolbox tackles this challenge by allowing users to easily test multiple circuit optimization techniques. 
For example, to reduce noise, a user could utilize a compiler that applies the ZX-calculus~\cite{kissinger2019reducing} to reduce the number of T-count gates in a circuit. 
However, the user could also apply zero-noise extrapolation~\cite{he2020resource}, an error mitigation technique that incrementally increases the number of C-NOT gates to predict the behavior of the circuit with zero noise.
Both optimization approaches aim to achieve the same goal while following two distinct approaches.
Some circuit optimization techniques might also require the intermediate execution of a circuit, for instance, to gather data about the performance of specific parameters, which is used by classical QAOA optimizers.

\subsubsection*{(2) Circuit Execution}

Quantum circuits are executed by selecting a matching backend. 
This can be a local simulator (with or without noise), an external simulator that uses a classical computing cluster, or a real quantum backend. 
Local simulation only requires the selection of a supported simulator, whereas execution on external backends usually requires the user to provide authentication, for instance, by inserting an API token.
The ProvideQ toolbox communicates with external backends via their API and currently supports IBM backends through Qiksit~\cite{Qiskit}, and Rigetti and IonQ backends through PlanQK~\cite{falkenthal_planqkplatform_2024}.
Users can decide on their own which concrete backend they want to use, but the toolbox can also assist them in their decision through the integration of automated backend selection tools such as the NISQAnalyzer~\cite{salm2020nisq} or the MQTPredictor~\cite{quetschlich2025mqt}.
Once a backend is selected, the ProvideQ toolbox checks again if the circuit matches the requirements of the backend. 
If this is the case, the circuit will be tanspiled and send to the backend. If this is not the case, an error message is returned and the user has to select another backend or recompile the circuit.

\subsection{Detailed API Specification}
\label{subsec:API_Specification}

This section describes the ProvideQ API in detail by explaining all exposed endpoints. 
Following the Meta-Solver strategy architecture, the API is structured around the concepts of Problem Types, Problem Instances, and Solvers.

The endpoints are grouped per Problem Type, each of which has a distinct set of endpoints where Problem Instances are referred to as \textit{/problems} and Solvers as \textit{/solvers}.
The interaction with the API is centered around Problem Instances, which can be created, read, and updated. 
Specifically, we define the following endpoints for every Problem Type:\\

\begin{tcolorbox}[colback=black!5, colframe=black!50, sharp corners, boxrule=0.2mm]
\small
\textbf{GET /problems/\{problemType\}}\\
Returns all Problem Instances of the given problemType.
\end{tcolorbox}

\begin{tcolorbox}[colback=black!5, colframe=black!50, sharp corners, boxrule=0.2mm]
\small
\textbf{GET /problems/\{problemType\}/\{problemId\}}\\
Returns all information of the Problem Instance associated with the given problemType and problemId.
\end{tcolorbox}

\begin{tcolorbox}[colback=black!5, colframe=black!50, sharp corners, boxrule=0.2mm]
\small
\textbf{POST /problems/\{problemType\}}\\
Creates a new Problem Instance of the given problemType.
Further information, such as the input of the problem, can be passed via the request body.
\end{tcolorbox}

\begin{tcolorbox}[colback=black!5, colframe=black!50, sharp corners, boxrule=0.2mm]
\small
\textbf{PATCH /problems/\{problemType\}/\{problemId\}}\\
The main request used to configure a solver. Updates the associated Problem Instance using the request body with the fields 'input', 'solverId', 'solverSettings', and 'state'.
To solve a Problem Instance, the 'state' field must be set to 'SOLVING', which requires all other fields to be set. 
Once the solving has been started, the Problem Instance can no longer be patched.
\end{tcolorbox}

\begin{tcolorbox}[colback=black!5, colframe=black!50, sharp corners, boxrule=0.2mm]
\small
\textbf{GET /problems/\{problemType\}/\{problemId\}/bound}\\
Returns a lower or upper bound for the possible solution of the associated Problem Instance.
\end{tcolorbox}

\begin{tcolorbox}[colback=black!5, colframe=black!50, sharp corners, boxrule=0.2mm]
\small
\textbf{GET /problems/\{problemType\}/\{problemId\}\\/bound/compare}\\
Returns a comparison of the solution and bound of the Problem Instance, requires those two values to be present.
\end{tcolorbox}

Endpoints for Solvers are similarly structured as the Problem Instance endpoints.
In our architecture, quantum and classical Solvers can be accessed through the same endpoints:

\begin{tcolorbox}[colback=black!5, colframe=black!50, sharp corners, boxrule=0.2mm]
\small
\textbf{GET /solvers/\{problemType\}}\\
Returns the name of all Solvers for the problemType.
\end{tcolorbox}

\begin{tcolorbox}[colback=black!5, colframe=black!50, sharp corners, boxrule=0.2mm]
\small
\textbf{GET /solvers/\{problemType\}/\{solverId\}/sub-routines}\\
Returns a list of subroutines that the associated Solver requires to be solved.
Solvers for the subroutines have to be selected later via PATCH requests of the Problem Instance.
\end{tcolorbox}

\begin{tcolorbox}[colback=black!5, colframe=black!50, sharp corners, boxrule=0.2mm]
\small
\textbf{GET /solvers/\{problemType\}/\{solverId\}/settings}\\
Returns a list of configurable settings for the given Solver.
\end{tcolorbox}

The differences between the quantum and classical solvers can be observed in their details, specifically in the defined subroutines. We use the definition of subroutines to handle the execution of quantum circuits by defining a special \textit{Quantum Circuit Processing} subroutine for every Solver that uses quantum computing. 
This approach allows us to model the Quantum Circuit Processing as a special Problem Type, and model the different compilers, backends, simulators, and error mitigators as Solvers for it. 
Thus, the API-based circuit execution is accessed in the same way as the solvers.

\section{Proof of Concept}

The goal of this section is to show that the ProvideQ toolbox is working as promised.
We demonstrate our concept by showing that the hybrid quantum-classical solvers from our running example (introduced in Figure~\ref{fig:vrp-running-example}) can be implemented as described. 
We configure both solvers via the ProvideQ HTTPS API, and evaluate their performance by letting them solve multiple VRP problems from the QOPTLib~\cite{osaba2023qoptlib} library, which is specifically designed for the evaluation of NISQ algorithms on combinatorial optimization algorithms.
We specifically note that the goal of this case study is just to show that the ProvideQ toolbox is capable of easily configuring hybrid solvers. 
Even though we will show results regarding the solvers' solution quality, our small case study does not allow any conclusion about their general performances.

\subsection{Experiment Setup}

We set up the experiment by implementing two VRP solvers based on the Meta-Solver Strategy introduced in Figure~\ref{fig:vrp-running-example}. We will call them the \textit{classical solver}, and the \textit{hybrid solver}.

The \textit{classical solver} tackles the VRPs using the Two-Phase Clustering approach by Laporte et al.~\cite{laporte2002classical}. 
The Two-Phase Clustering divides the VRP solving into a clustering and a routing phase. 
After selecting this solving method, the ProvideQ toolbox solves the clustering phase with a classical knapsack solver, decomposing the initial VRP problem into multiple TSP instances. 
Afterwards, we solve the TSP instances with the classical state-of-the-art solver LKH-3\cite{helsgaun2017extension}. 
The ProvideQ toolbox then interprets the TSP results and composes a solution to the intial VRP.

The \textit{hybrid solver} tackles the VRPs similarly.
It also uses the Two-Phase Clustering, and the same classical knapsack solver for the clustering phase.
Thus, differences between both the classical and hybrid solver are only made in the routing phase. 
The hybrid solver tackles the routing phase, meaning the solving of the TSP instances, by transforming them into QUBOs which are then solved with quantum annealing. 
The ProvideQ toolbox handles the QUBO transformation using the well-known encoding from Lucas~\cite{lucas2014ising}. For the quantum annealing, we decided to use the D-Wave state vector simulator.
After setting up the solvers, we run and compare them on all VRP problems from the QOPTLib~\cite{osaba2023qoptlib}. 
QOPTLib includes a total of ten VRP instances with a size ranging from 4 to 8 citites. 
To make the problems solvable for small-scale NISQ devices, QOPTLib also adds additional simplifications to the VRP problems. For example, setting the demand for every city to one and ensuring that every VRP can be solved with a maximum of two routes.
We give every solver ten tries for every problem, concluding a total of 100 runs. The code to reproduce this experiment is available on GitHub: 
\href{https://github.com/ProvideQ/provideq-tool-paper-poc}
{https://github.com/ProvideQ/provideq-tool-paper-poc}.

\subsection{Results}
\begin{figure*}[ht]
    \centering
    \includegraphics[width=\textwidth]{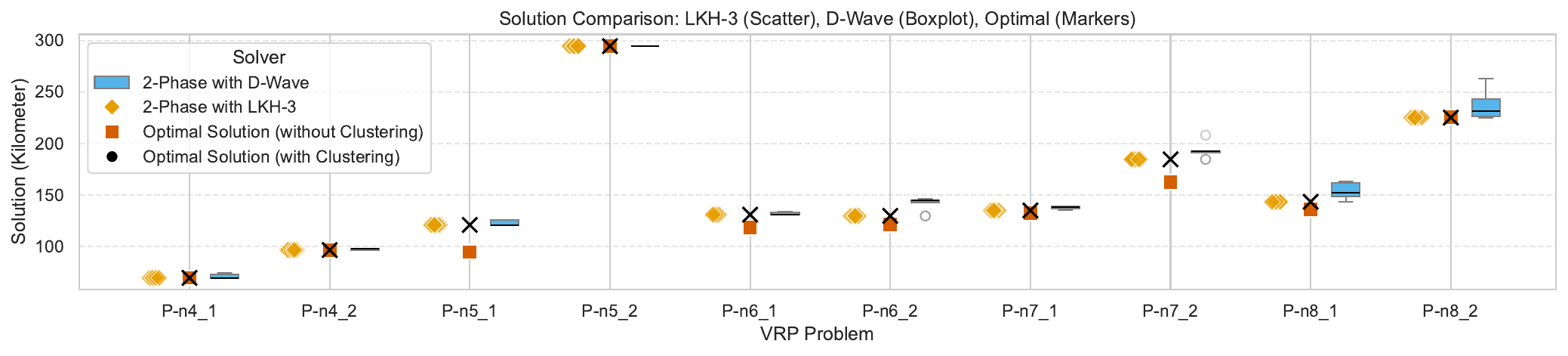}
    \caption{Results from the VRP Solving case study.}
    \label{fig:case_study}
\end{figure*}

The results of our experiment are shown in Figure~\ref{fig:case_study}, plotting the names of the QOPTLib VRP problems on the x-axis and the solution (length of the route in kilometers) on the y-axis. 
Four different types of entries are plotted for each problem, including two different types of optimal solutions.
The optimal solution (without clustering) shows the best possible solution for the given VRP problem, whereas the optimal solution (with clustering) shows the best possible solutions that our solvers are able to reach.
Our solvers are not always able to reach a global optimum because when we apply the clustering, we sometimes lose the opportunity to take specific routes.

Analyzing the solutions of the solvers, we observe that the classical solver was always able to compute the best possible solution in every attempt. 
Compared to this, the solutions of the hybrid solver scatter. 
It was also able to find a valid solution in all attempts, but the best possible solution was only found in some. 
These observations are as expected, as the problems included in the QOPTLib benchmarking set are very small and do not pose any challenge to the classical solver. 
They are, however, challenging for quantum-based approaches, especially because we are reliant on noisy hardware or non-scalable simulations.
Furthermore, the hybrid solver solves the TSPs with a QUBO transformation, which unavoidably applies approximations through the introduction of penalized slack variables. 
The hybrid solver also utilizes a quantum annealer that uses a generalized approximation approach and, compared to LKH-3, is not specially optimized for TSPs.
Given the observations of this case study, we can conclude that the classical solver is superior for the small-scale problems included in the QOPTLib.
The hybrid solver, even with the additional approximations and simulation overhead, is able to produce reasonable results, with the chance of finding an optimal solution in some attempts.

While observing the solver's performances we also measured their runtimes. 
The total runtime (including clustering, routing, and the composition of the final result) always required less than 1 second for the classical solver and between 2 to 8 seconds for the hybrid solver.
These numbers also include all overheads that the ProvideQ toolbox adds, for instance, when parsing inputs or calling solvers.
While those numbers are interesting to observe, we want to state that a direct comparison of runtimes is not meaningful in this context. 
This is because the used state vector simulator simulates the behavior of a quantum annealer, and thus introduces an additional computational overhead that would not occur on real hardware.

\section{Related Work}

Quantum computing solutions for optimization problems are highly anticipated and thus a lot of research initiatives were created to investigate their performance and build platforms, frameworks, and libraries to use them.

The PlanQK platform~\cite{falkenthal_planqkplatform_2024} is designed as a collaborative space to share quantum back-end adapters, quantum/hybrid algorithm implementations, and quantum workflows, and integrates a visual workflow editor to compose quantum services for higher-order applications.
Similarly to ProvideQ, users can easily access different optimization algorithms via an API and integrate them into their external tools.
The MQT Quantum Auto Optimizer~\cite{volpe2024predictive}, Quafu-Qcover~\cite{xu2024quafu}, QPLEX~\cite{qplex}, or PyQUBO~\cite{pyQubo} are frameworks that enable easy access to quantum optimization solutions by implementing strategies to automatically translate optimization models into quantum formulations. They enable users to solve these formulations with established quantum algorithms such as QAOA, VQE, or Grover, and execute them on the backends of large vendors.

ProvideQ differentiates itself from the mentioned platforms and frameworks by focusing mainly on the implementation of Meta-Solver strategies.
We implement techniques to decompose large optimization models and solve them by combining several classical and quantum subroutines, instead of just translating problems and then applying a single algorithm.
Further, we anticipate the good performance of existing classical solvers and enable users to use them in combination with quantum subroutines, instead of replacing them. 

Approaches that have great similarities to the Meta-Solver strategy approach are implemented in the Quokka framework~\cite{beisel2022quokka}, which enables the service-based execution of quantum algorithms with a workflow model~\cite{beisel2024utilizing}, and the QuaST decision tree~\cite{poggel2024creating}, which automated the execution of quantum solutions with a decision diagram. 
Both of these frameworks implement technologies that allow users to impact the execution process by configuring its details via well-defined interfaces. 
The ProvideQ toolbox implements similar techniques but is again unique due to focus on hybrid quantum-classical solvers that extend classical state-of-the-art solvers with quantum subroutines. 

\section{Conclusion and Future Work}

We presented the ProvideQ toolbox, which makes hybrid quantum-classical solutions for combinatorial optimization problems accessible to a broad audience by providing a software stack that combines methodologies from polylithic mathematical modeling, classical and quantum combinatorial optimization, as well as quantum circuit compilation and deployment.
Users benefit from the ProvideQ toolbox by gaining the opportunity to explore new hybrid quantum-classical optimization techniques via our Frontend, or integrate them into their external software using our API.

This paper shows that hybrid quantum-classical optimization is already usable today and that the quantum software stack is mature enough to support the implementation of such a sophisticated tool.
However, we also anticipate that quantum technologies are not yet mature enough to compete with highly optimized classical technologies for industry-relevant use cases such as the vehicle routing problem.
Quantum speedups for practical applications are still a concept of the future and require consistent improvements in quantum hardware and algorithms.

The development of the ProvideQ toolbox is an ongoing initiative. In future work, we will continue to integrate more Meta-Solver strategies. This includes decomposition techniques and algorithms, as well as the integration of more quantum circuit compilation tools. 
Furthermore, we aim to investigate new methods to predict the resource consumption and solution quality of classical and quantum solvers to provide a guidance framework for the creation of good solvers.

\section*{Acknowledgments}

The authors thank Tobias Czetsch, Piotr Malkowski, and Luke Southall for helping us with the implementation of the ProvideQ toolbox.
We acknowledge that we used the \textit{Adobe Firefly} AI to generate the icons used in Figure~\ref{fig:user_interaction}. 

\balance

\bibliographystyle{IEEEtran}
\bibliography{provideq}

\end{document}